\documentclass[aps,pre,twocolumn,showpacs,amssymb]{revtex4}
\usepackage[dvips]{graphicx}
\usepackage{dcolumn}
\usepackage{bm}

\def\gs{\gtrsim}
\def\ls{\lesssim}
\def\be{\begin{equation}}
\def\en{\end{equation}}    
\def\gs{\gtrsim}
\def\ls{\lesssim}

\newcommand{\bi}[1]{\mbox{\boldmath$#1$}}
\newcommand{\av}[1]{\langle{#1}\rangle}
\newcommand{\AV}[1]{\bigg\langle{#1}\bigg\rangle}
\def\p{\partial}
\def\bea{\begin{eqnarray}}
\def\ena{\end{eqnarray}}

\renewcommand{\theequation}{\arabic{section}.\arabic{equation}}

\begin{document}
\draft
\bibliographystyle{prsty}
\title{Phase separation 
 in a   binary mixture confined 
between   symmetric parallel plates: Capillary condensation transition 
near the bulk critical point}
\author{Syunsuke Yabunaka$^1$, 
Ryuichi Okamoto$^2$,   and Akira Onuki$^1$}
\address{$^1$Department of Physics, Kyoto University, Kyoto 606-8502\\
$^2$Fukui Institute for Fundamental Chemistry, Kyoto University, Kyoto 606-8103}
\date{\today}

\begin{abstract} 
We investigate   phase separation of near-critical binary mixtures 
 between parallel symmetric walls 
in the strong adsorption regime.   
We take into account the 
renormalization effect due to  the critical fluctuations 
using the recent local functional theory 
$[$J. Chem. Phys. {\bf 136}, 114704 (2012)$]$.
In statics, a  van der Waals loop is obtained 
in the relation between the average 
order parameter $\av{\psi}$ in the film and the chemical potential 
when  the  temperature $T$ is lower than the film critical temperature 
$T_c^{\rm ca}$ (in the case of  
an upper  critical solution temperature).  In dynamics, we lower $T$ 
below the    capillary condensation line  from   above $T_c^{\rm ca}$.   
We calculate   the subsequent time-development 
assuming no  mass exchange between the film and the reservoir. 
In the early stage,  the order parameter $\psi$  changes 
only in the direction  perpendicular to the walls. 
For sufficiently deep quenching, such one-dimensional 
profiles  become unstable with respect to the 
fluctuations varying in the lateral directions. 
The late-stage   coarsening  is then accelerated 
by the hydrodynamic interaction. 
 A  pancake domain of the phase disfavored by the walls  
 finally appears in the middle of the film. 
 \end{abstract}

\maketitle

\pagestyle{empty}

\section{Introduction}
\setcounter{equation}{0}
 
The  phase behavior of fluids  
confined  in narrow regions 
has been studied extensively  
\cite{Evansreview,Gelb,Binder}.
It strongly depends on the geometry of the walls and 
on the molecular interactions 
between the fluid and the  walls.
Its understanding is crucial
 in the physics of fluids in porous media. 
It  is also needed to study the dynamics of confined fluids.  

In particular,  the liquid phase is usually 
favored by the walls in one-component fluids, 
while one    component  
is preferentially attracted 
to the walls in  binary mixtures  
\cite{Binderreview,PG,Cahn,Bonn,Lawreview,Rudnick,Liu}. 
In such situations,  narrow regions may be  
filled with the phase favored by  the  
walls or may hold some fraction of 
the disfavored  phase. 
Hence, in the film geometry, 
there appears   a first-order  phase transition 
  between these states, which forms 
a line (CCL) 
ending  at a film critical point  outside the bulk coexistence curve  
  in the  $T$-$\mu_\infty$ plane  
\cite{Evansreview,Gelb,Binder,Binder-Landau,Nakanishi,Evans-Marconi,p}, 
where  $\mu_\infty$ is 
 the reservoir chemical potential \cite{chemical}.     
We call it the capillary condensation transition even for binary mixtures,  
though this name has been  used for the gas-liquid phase 
transition  in porous media  \cite{Gelb}. 
Around  CCL,  the reservoir is rich in 
the component  disfavored by the walls  for binary mixtures.
With increasing  the  wall separation $D$, 
the film critical point  approaches the bulk 
critical point. Crossover then occurs    
between   two-dimensional  (2d)  
and  three-dimensional  (3d) phase transition behaviors.

 For Ising films near the bulk criticality,    
Fisher and Nakanishi \cite{Nakanishi} 
presented the scaling theory 
of CCL  in the $T$-$h$ plane, 
where $h$  represents  applied magnetic field.  
 They also  calculated CCL in the mean-field $\phi^4$ theory. 
Evans {\it et al}. used the density functional theory 
to calculate the inhomogeneous structures in pores \cite{Evans-Marconi}. 
For a Lennard-Jones fluid in cylindrical pores, 
Peterson {\it et al.}\cite{p}  obtained  steady gas-liquid two-phase   patterns. For a lattice gas model, Binder and Landau \cite{Binder-Landau} 
studied the capillary condensation transition using a Monte Carlo method. 
For a microscopic model of 2d  Ising stripes, 
 Macio\l ek {\it et al} \cite{Evans-Anna} 
 found a (pseudo) CCL using a density-matrix renormalization-group  method.  
For square well  fluids in slit pores, 
Singh {\it et al.}\cite{Singh}  
numerically examined the crossover from 3d to 2d.

Recently, two of  the present authors \cite{Oka} 
calculated CCL near the bulk critical point 
 using the local functional theory \cite{Yang,Upton},   
which accounts for the renormalization effect 
due to the  critical fluctuations. The lowering 
of the film critical temperature $T_c^{\rm ca}$ from 
the bulk critical temperature $T_c$  
was shown to be proportional to 
 $D^{-1/\nu}$ (where $\nu\cong 0.63$) in accord with 
the scaling theory \cite{Nakanishi}. 
 Along CCL, our calculations \cite{Oka} 
and those by Macio\l ek {\it et al} \cite{Evans-Anna} 
 showed   strong enhancement 
 of  the so-called Casimir amplitudes \cite{Casimir}. 
Similar first-order transitions  were found 
 between  plates  \cite{Tsori}  and colloids \cite{Okamoto} 
 in  binary mixtures 
containing salt.

The  aim of this paper is  to investigate 
the phase separation  in near-critical binary mixtures  
 between parallel plates  
  using  model H and model B \cite{H77,Onukibook}. Here,  
 phase separation  takes  place 
around  CCL and   the hydrodynamic interaction is crucial  in 
the late-stage phase separation. 
It is worth noting that near-critical fluids in porous media  exhibit 
history-dependent frozen domains and 
 activated dynamics with non-exponential relaxations 
\cite{Goldburg,Wil}. To gain insight into  such 
complicated effects, we  may start with  near-critical fluids 
in  the   film geometry.  
Treating near-critical fluids,  we may  construct a universal 
theory  with  a few materials-independent  parameters, where 
 $D$ much exceeds  microscopic spatial scales.

In the literature, much attention has been paid to 
the interplay of wetting and phase separation 
\cite{Tanakareview,Liu-Durian,Bates,Das}, 
which  is referred to as 
 surface-directed phase separation. However, 
simulations including  the hydrodynamic interaction 
have not been  abundant  
\cite{Tanakareview,Araki,Yeomans,Puri}. 
We mention that Tanaka and Araki \cite{Araki} 
integrated  the model H equations 
in the semi-infinite situation and  Jaiswal {\it et al.}
\cite{Puri} performed 
 molecular dynamics simulation to 
investigate the hydrodynamic flow effect  between parallel plates. 
In our simulation,  the order parameter $\psi$ 
changes  in the direction perpendicular  to  the walls 
in  the strong adsorption regime. 
Then, the dynamics is one-dimensional 
in an   early stage  but the fluid flow  in the lateral 
directions accelerates the  late-stage coarsening 
even under the no-slip boundary condition on the walls 
\cite{Tanakareview,Yeomans,Puri,Araki,Onukibook}.  

On the other hand, 
Porcheron and Monson \cite{Monson} numerically 
studied the dynamics of extrusion 
and intrusion of liquid mercury 
between a cylindrical pore   and a  reservoir. 
Such a process  is crucial 
in experiments of adsorption and desorption between  a  porous material  
 and a surrounding fluid \cite{Gelb}. 
In our simulation we assume no  mass exchange 
imposing the periodic boundary condition 
in the lateral directions,  as in the previous  simulations of 
 surface-directed phase separation.

The organization of this paper is as follows. 
In Sec.II,  we will summarize the results 
of the   local functional theory 
 of   near-critical binary mixtures in the film geometry. 
We will newly present   some   results 
on the phase behavior, which will facilitate 
 understanding  the  
phase separation near  CCL. 
In Sec.III, we will present our simulation results 
of the phase separation with the velocity field (model H) 
and without it (model B).

\setcounter{equation}{0}

\section{Theoretical background}

This section provides   the theoretical background of our simulation 
on the basis of our previous paper \cite{Oka}.
The Boltzmann constant $k_B$  will be set equal to unity.

\subsection{Ginzburg-Landau  free energy}

We suppose near-critical binary mixtures with  an   
upper  critical solution temperature (UCST)  $T_c$ 
at a constant  pressure. The order parameter $\psi$ 
is proportional to $c-c_c$, 
where $c$ is the composition and $c_c$ is its critical value. 
The reduced temperature is written as 
\be 
\tau= (T-T_c)/T_c.
\en 
In our numerical analysis, the usual critical exponents  
 take the following values\cite{Onukibook}: 
\bea 
&&
\alpha=0.110, \quad \beta=0.325,  \quad \gamma=1.240, \nonumber\\
&& \nu=0.630,  \quad \eta=0.0317,\quad \delta=4.815.  
\ena 
At the critical composition with  $\tau>0$, 
the correlation length 
is written as  $\xi = \xi_0 \tau^{-\nu}$, 
where $\xi_0$ is a microscopic length.  
 The coexistence curve in the 
region  $\tau<0$ 
is  denoted by  CX.  The correlation length on CX is written as 
$\xi=  \xi_0' |\tau|^{-\nu}$, where $\xi_0'$ is 
 another microscopic length   
 with the ratio $R_\xi= \xi_0/\xi_0'$ 
being a universal number. We write $\psi$  
 in the coexisting two phases as $\pm \psi_{\rm cx}$ with 
\be 
\psi_{\rm cx} = b_{\rm cx}|\tau|^\beta,
\en  
where  $ b_{\rm cx}$ is a constant.

We assume that   the  bulk free energy $F$ 
including the gradient part is of the local functional form 
\cite{Yang,Upton,Oka},   
\be
F  = \int d{\bi r}[f + \frac{1}{2}T_c C|\nabla\psi|^2].  
\en 
In the following, we give a simple  form for 
 the free energy density  $f= f(\psi,\tau)$.  
In our theory,  the critical fluctuations with sizes 
smaller  than the  correlation length $\xi$ have already been 
coarse-grained at the starting point.

\subsection{Coexistence-curve exterior}

Outside CX,  $f$  is of the  Ginzburg-Landau form,   
\be
{f}/{T_c}=  \frac{r}{2}\psi^2+ \frac{u}{4}\psi^4.
\en 
Here,  we have omitted the free energy contribution 
for $\psi=0$, whose singular part is proportional 
to $|\tau|^{2-\alpha}$  yielding the specific heat singularity. 
The coefficients   $r$ and $u$ in $f$ and  $C$ in $F$  are 
 renormalized ones  in three dimensions.   As in 
the linear parametric model \cite{Sc69},  we use  
a  nonnegative parameter  $w $ representing   the distance from the 
critical point in the $\tau$-$\psi$ plane  
to obtain  
\bea 
r/\tau&=& C_1 \xi_0^{-2}w^{\gamma-1}  ,\\
u/u^*&=&  C_1^2  \xi_0^{-1}w^{(1-2\eta)\nu}, \\
C& =& C_1 w^{-\eta\nu}, 
\ena
where   $C_1$ and $u^*$ are constants. 
We may  set  $C_1=1$ 
 by rescaling $ C_1^{1/2}\psi \to \psi$ without loss of generality.  
In the present case, $(C_1\xi_0)^{1/2}\psi$ is dimensionless.  
The constant $u^*$ is a universal number  
and we  set  $u^*= 2\pi^2/9$. 
The fractional powers  of $w$ in  
Eqs.(2.6)-(2.8) arise from the renormalization 
of the critical fluctuations with wavenumbers larger than the 
inverse correlation length $\xi^{-1}$. 
We determine  $w$ as a function of 
$\tau$ and $\psi$  by  
\be 
w=  \tau +    (3u^* C_1\xi_0) w^{1-2\beta}\psi^2,      
\en 
which is equivalent to  $w^\gamma= (r+3u\psi^2)\xi_0^{2}/ C_1$. 
Thus, $w=\tau$ for $\psi=0$ and $\tau\ge 0$, while 
 $|\psi|\propto w^{\beta}$ for $\tau=0$. 

The derivative $\mu=\p f/\p \psi$ at fixed $\tau$ 
denotes  the   chemical potential 
difference between the two components\cite{chemical,Yang,Liu}, 
but it will be  simply called 
the chemical potential.  In terms of the ratio  $S=\tau/w$,  
it reads  
\be 
\frac{{\mu}}{T_c}
 = \frac{2-\alpha+ 4(1-\alpha)S +5\alpha S^2 }{6[2\beta 
 +(1-2\beta)S]\xi_0^{2}} C_1 w^\gamma  \psi.   
\en
 On  CX, we require  $\mu=0$, which yields the equation  
 $2-\alpha+ 4(1-\alpha)S +5\alpha S^2=0$ for $S$. 
On CX, this gives  
 $S = -1/\sigma$ or $w= -\sigma \tau$ 
with $\sigma=1.714$. Together with   Eqs.(2.3) and (2.9),  
we obtain   
\be 
b_{\rm cx}^2= (1+\sigma)\sigma^{2\beta-1}/3u^*C_1\xi_0.
\en      
W introduce  the susceptibility $\chi=\chi(\tau,\psi)$  defined by 
\be 
T_c/\chi= \p \mu/\p \psi= \p^2 f/\p \psi^2 .
\en 
 For  $ \psi=0$ and $\tau>0$, we simply obtain  
$\chi(\tau,0)=C_1^{-1} \xi_0^2 \tau^{-\gamma}$. On CX,  
we write $\chi_{\rm cx}= \chi(\tau,\psi_{\rm cx})$. 
In terms of  the critical amplitude ratio $R_\chi= \chi(|\tau|,0)
/\chi_{\rm cx}$ for $\tau<0$, the susceptibility on CX reads     
\be 
\chi_{\rm cx}= R_\chi^{-1}   C_1^{-1} 
\xi_0^2 |\tau|^{-\gamma}.   
\en 
Some calculations give $R_\chi= 8.82$ \cite{comment0}. 
In terms of $\chi$,  the correlation length 
is expressed as  $\xi=(C\chi)^{1/2}$, which yields  
the critical amplitude ratio $R_\xi= \xi_0/\xi_0'=2.99$  \cite{comment0}.  
For $\tau=0$, we have $\xi \propto |\psi|^{-\nu/\beta}$.

\subsection{Coexistence-curve interior}

The interior of  CX is given by  
$|\psi|<\psi_{\rm cx}$ and $\tau<0$, where we need 
to define the free energy density 
$f$ to examine two-phase coexistence.   
We  assume 
a  $\psi^4$-theory with  coefficients depending  only on $\tau$, where  
$\p f/\p \psi=\mu$ and $ \p^2 f/\p \psi^2= T_c/\chi$ are continuous 
across  the coexistence curve. We then obtain  
\be 
({f}-{f_{\rm cx}})/ {T_c}= ({ \psi_{\rm cx}^2 }/{8 \chi_{\rm cx}}) 
(\psi^2/\psi_{\rm cx}^2-1)^2  ,
\en 
where $f_{\rm cx}$ 
is the free energy density  on CX  and 
$\chi_{\rm cx}$ is defined by Eq.(2.13).  
We also set  
\be 
C=C_{\rm cx}=   C_1 |\sigma\tau|^{-\eta\nu},
\en 
 which is the 
value of $C$ on CX.  The renormalization effect 
inside CX  is assumed to be unchanged from   that  on CX with the same $\tau$. 
The   ${\mu}$ inside CX then reads  
\be 
\mu /{T_c}=  
(\psi^2/\psi_{\rm cx}^2-1)\psi/{2 \chi_{\rm cx}}   
\en 

The surface tension $\sigma$ between coexisting bulk two phases 
is given by the standard expression, 
\bea 
\sigma&=& 2T_c (C_{\rm cx}/\chi_{\rm cx})^{1/2}\psi_{\rm cx}^2/3\nonumber\\ 
&=& A_s  T_c /\xi^2, 
\ena  
where $\xi=\xi_0'|\tau|^{-\nu}$ is the correlation length on CX. 
The  universal number $A_s$ is 
estimated to be  $0.075$ in our model, while its 
 reliable value is about $0.09$ \cite{Oka}.

\subsection{Near-critical fluids between parallel plates}

\begin{figure}[t]
\begin{center}
\includegraphics[scale=0.4]{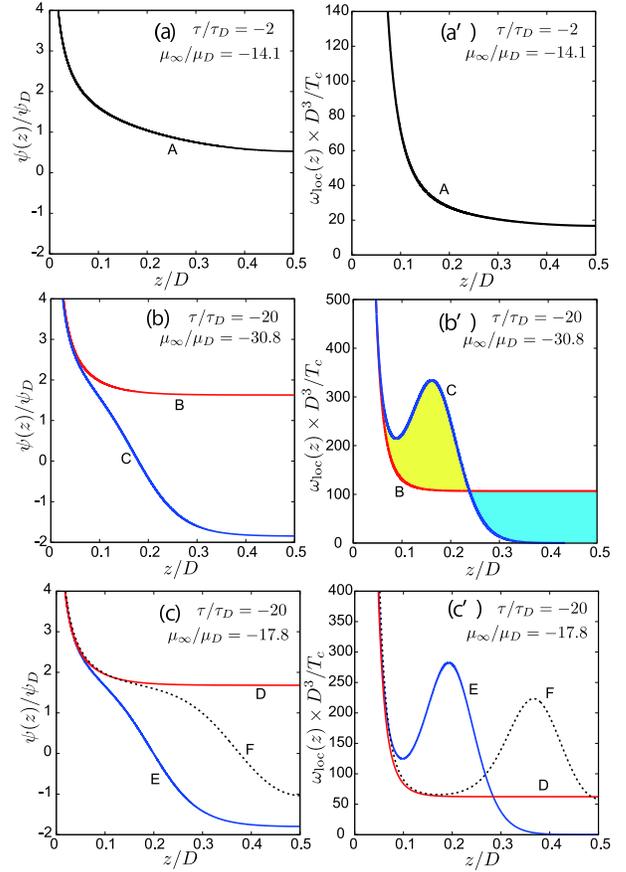}
\caption{(Color online) 
Normalized 1d  profiles 
$\psi(z)/\psi_D$ (left) 
and $\omega_{\rm{loc}}(z)D^3/T_c$ (right) vs $z/D$ 
for $(\tau/\tau_D, \mu_\infty/\mu_D)
=(-2, -14.1)$ (top), 
 $(-20, -30.8)$ (middle), and $(-20, -17.8)$ (bottom). 
Top: Adsorption-dominated profile  A with $\psi(z)>0$. 
Middle:  Two profiles B  and C on  the capillary condensation line  
with the same grand potential $\Omega$.  In (b'), the two curves of 
  $\omega_{\rm{loc}}(z)$ enclose 
two regions with  the same area close to the surface tension $\sigma$. 
 Bottom:  
Three profiles D,E, and F with   
the same $\tau$ and  $\mu_\infty$ (see Fig.3). In  (c'), 
$\Omega$ increases in  the order  of D, E, and F.   
}
\end{center}
\end{figure}

\begin{figure}[t]
\begin{center}
\includegraphics[scale=0.6]{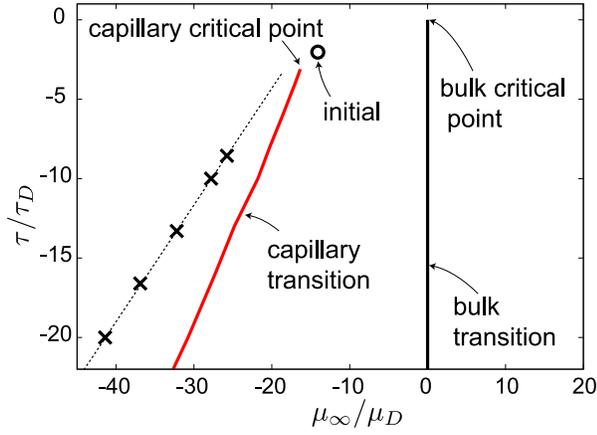}
\caption{(Color online) 
Phase diagram   of a near-critical 
fluid  in a film for large adsorption in the $\mu_\infty/\mu_D$-
$\tau/\tau_D$ plane.   The bulk 
coexistence line is given by $\tau<0$ and $\mu_\infty=0$. 
On its left,   
there appears a first-order capillary condensation  line (red bold line) 
ending at  a film critical point, 
which is calculated from 1d profiles. Displayed 
also are   values of $\mu_\infty/\mu_D$ 
in steady two-phase coexistence in our simulation of a $2D\times 2D\times D$ 
system ($\times$). Starting  point of our simulation ($t<0$)  
is also shown ($\circ$). 
}
\end{center}
\end{figure}

\begin{figure}[t]
\begin{center}
\includegraphics[scale=0.55]{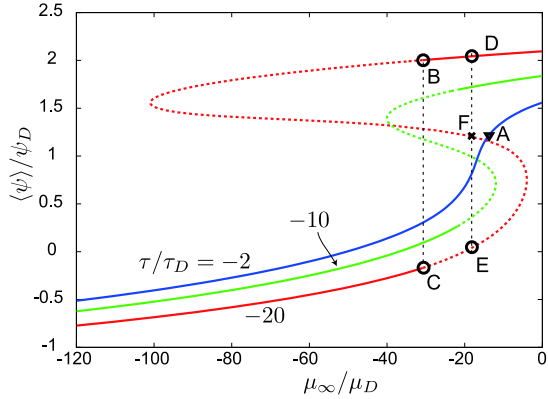}
\caption{\protect
(Color online) Isothermal curves in the  
$\mu_\infty/\mu_D$-$\av{\psi}/\psi_D$ plane,
which  are calculated for 1d profiles at $\tau/\tau_D=-2$, -10,  and -20. 
For $\tau$ less  than its film critical 
value ($= -3.14 \tau_D$),  a van der Waals loop appears. 
Dotted parts of the  curves for  $\tau/\tau_D=-10$ and -20 
are not stable in contact with a reservoir.   
Points A,B,C,D,E, and F correspond to the curves in Fig.1.}
\end{center}
\end{figure}

We  suppose  
 a near-critical fluid     between parallel symmetric walls 
 in the region $0<z<D$, where $D$ is much longer than 
any microscopic lengths. To  avoid the discussion of the edge effect, 
  the   lateral plate dimension  $L$ is supposed to much exceed $D$. 
The fluid is close to the bulk criticality and above 
the prewetting transition line \cite{Evansreview,Cahn,PG,Bonn}.  
We use our  local functional 
theory, neglecting 
the two-dimensional thermal  fluctuations with sizes  exceeding $D$ 
in the $xy$ plane.

We  scale $\tau$ and   $\psi$  
in units of  $\tau_D\propto D^{-1/\nu}$ 
and $\psi_D \propto D^{-\beta/\nu}$, respectively,  defined by 
\bea 
\tau_D &=& (\xi_0/D)^{1/\nu}, \\
\psi_D &=& (24^{\beta/\nu} /3u^* C_1 \xi_0)^{1/2} \tau_D^{\beta}.  
\ena 
In equilibrium theory,  it is convenient to assume that 
the fluid between the walls  is   
 in contact with  a large reservoir   
 containing the same binary mixture, where   the  order parameter 
 is $\psi_\infty$ and   the chemical potential   is 
\be 
\mu_\infty=\mu(\psi_\infty, \tau).  
\en  
Here,   $\mu_\infty$ 
corresponds to  magnetic field $h$ for films of 
Ising spin systems. We are interested in 
the case  $\mu_\infty<0$ (or $\psi_\infty<0$)  and $\psi_0>0$, 
where $\psi_0$ is the  value of $\psi$ at the walls. 
If equilibrium is attained in the total system including the reservoir, 
we should minimize   the  film grand potential $\Omega$. 
 Including the surface free energy, we assume the form, 
\be 
\Omega= \int d{\bi r}~\omega_{\rm{loc}} 
- T_c \int  dS ~h_1\psi ,
\en 
where the  space integral $\int d{\bi r}$   is 
within the film,    the surface integral $\int dS$  
is on the walls  at $z=0$  and $D$, 
and  $h_1$ is a   surface field symmetrically  given  on the two walls. 
In addition, we neglect the surface free energy  
 of the form  $\int dS \lambda^{-1} \psi^2$    
assumed in the literature\cite{Cahn,Bonn,Binderreview,PG} 
(or we consider the limit $\lambda \to \infty$). 

In Eq.(2.21) 
 $\omega_{\rm{loc}}$ is the local grand potential density including the 
gradient part,  
\be 
\omega_{\rm{loc}}= \omega_s + \frac{1}{2}T_cC |\nabla\psi|^2, 
\en 
where   $\omega_s $ is the excess grand potential 
density written as 
\be
\omega_s  =  f(\psi)-f(\psi_\infty) 
 - \mu_\infty (\psi-\psi_\infty). 
\en  
Now minimization of $\Omega$ yields  the bulk equation, 
\be 
\frac{\delta F}{\delta \psi}= 
\mu - \frac{T_c}{2}C'|\nabla\psi|^2 -T_c 
C \nabla^2\psi=\mu_\infty,  
\en 
where $C'= \p C/\p\psi$.  The boundary conditions   
  at $z=0$ and $D$ are   given by  
\be 
\psi'(x,y,0)=- \psi'(x,y,D)=   -h_1/C.     
\en  
where $\psi' = \p \psi/\p z$. 

The  role of $h_1$ in this paper 
is simply to assure  the strong adsorption regime 
$\psi_0/|\tau|^\beta \gg (C_1\xi_0)^{-1/2}$ \cite{Liu,Yang,Oka},  
where $\psi_0$ is the boundary value of $\psi$. 
This regime  is eventually realized on approaching  the criticality 
(however small $h_1$ is). 
In our  simulation,  the profile of $\psi$ in 
the region $0<z<\xi$ is nearly one-dimensional depending only on $z$ 
 even in two phase states (see Figs.5 and 6). It  decays 
  slowly   as  $ (\ell_0+z)^{-\beta/\nu}$ for $0<z<\xi$ 
 \cite{Liu,Lawreview,Rudnick,Oka}, where    $\ell_0$  
is a short microscopic length introduced by Rudnick and Jasnow \cite{Rudnick}. 
With the gradient free energy in the form of Eq.(2.22),  $\ell_0$
 is expressed as  
\be 
\ell_0 \sim \xi_0 (C_1\xi_0)^{-\nu/2\beta} \psi_0^{-\nu/\beta} 
\sim D (\psi_D/\psi_0)^{-\nu/\beta},
\en   
in terms of $\psi_0$.  
The excess surface adsorption of $\psi$ in the region $0<z<\ell_0$ 
is of order  $\psi_0\ell_0 \sim \psi_0^{1-\nu/\beta}$ 
and is  negligible for large $\psi_0$ from  $\beta/\nu \sim 2$, 
while that  in the region $\ell_0 <z<\xi$ 
is of order    $\xi^{1-\beta/\nu}$ for $\xi(\tau,\psi_m) <D/2$ 
 \cite{Liu,Rudnick,Lawreview}. 
In the strong-adsorption regime we 
calculate   the  average of $\psi$ along the $z$ axis,  
\be 
\av{\psi}= \int_0^D  dz \psi/D.    
\en 
 In two-phase states,  
$\av{\psi}$  depends on $(x,y)$.

From   Eq.(2.25) it follows the estimation   
$h_1 \sim C(\psi_0) \psi_0/\ell_0 $.  
As $h_1/ |\tau|^{\beta\delta-\nu}  \to \infty$, 
we  find      
\be 
 h_1 = \psi_0^{\delta-\nu/\beta}(B_1 + B_2 \tau\psi_0^{-1/\beta}+\cdots), 
\en 
where     $B_1$ and $B_2$ are  positive  constants. 
This is the expression for $D \to \infty$. 
In this regime, 
 the surface free energy  in Eq.(2.21) is  given by  
 $- 2B_1\psi_0^{2\nu/\beta}A$, where $A=L^2$ is the surface area. 
In our previous paper\cite{Oka}, 
we  examined  the film phase behavior  
at fixed large $\psi_0$, 
 treating the surface free energy as a constant. 
In our  simulation,  
we assume  the boundary 
condition (2.25) with $h_1/C=1011\psi_D/D$  
to obtain  $\psi_0/\psi_D \cong14.8$.

\subsection{ Capillary condensation  transition} 

We consider the capillary condensation transition 
on the basis of one-dimensional (1d)  profiles $\psi=\psi(z)$.  
From Eq.(2.25), we have 
the symmetry $\psi(z)= \psi(D-z)$. 
In the region $0<z<D/2$, Eq.(2.24) is integrated to give   
 \be 
z= \int^{\psi_0}_\psi d\psi 
\bigg[\frac{C(\psi)/2}{\omega_s(\psi)+\Pi}\bigg]^{1/2}. 
\en 
Here, $\Pi= -{A^{-1}}{\p \Omega}/{\p D}$ is the osmotic pressure.   
It   is    the force density per unit area 
 exerted by the fluid  to the plates. In our case, $\Pi<0$,  
 indicating    attractive inter-wall interaction.  In the 1d  case,  
it is also written as    
\bea 
\Pi&=&  f(\psi_\infty)-f(\psi_m)- 
 \mu_\infty (\psi_\infty-\psi_m) \nonumber\\
&=& -\omega_s(\psi_m).
\ena  
At the midpoint  $z=D/2$, we set  $\psi_m= \psi(D/2)$.  
The fluid at the midpoint  
can be  in the phase favored by the walls 
with $\psi_m \sim \psi_{\rm cx}$  due to the strong adsorption 
on  the walls 
or in the disfavored phase  with $\psi_m \cong \psi_{\infty}<0$.  
Equation (2.30) indicates  
  $\Pi\cong  2\mu_{\infty} \psi_{\rm cx}$ in the 
former case and 
 $\Pi \cong  -T_c( \psi_m-\psi_\infty)^2/2\chi_{\rm cx} 
\cong 0$  in the latter case, so $\Pi$ can be very different in 
 these two cases \cite{comment2}.

\begin{figure}[t]
\begin{center}
\includegraphics[scale=0.55]{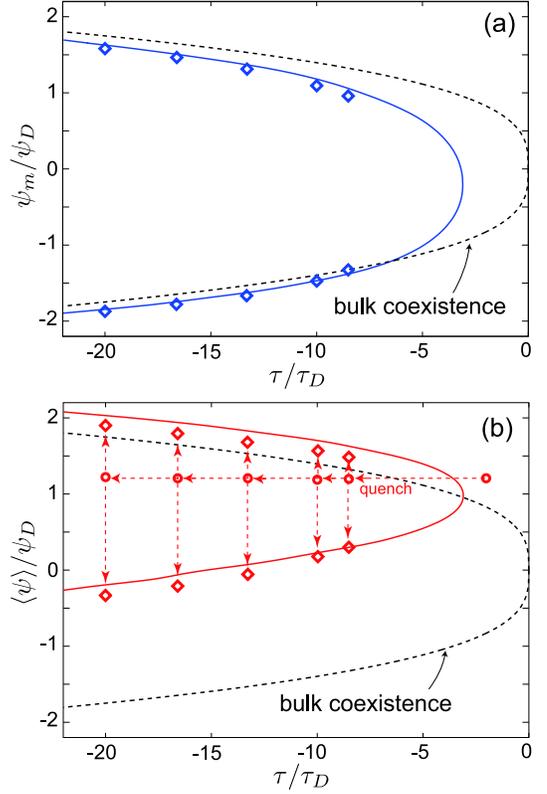}
\caption{\protect
(Color online)  Phase diagrams   in the 
$\tau/\tau_D$-${\psi_m}/\psi_D$ plane  in  (a) and  in the  
$\tau/\tau_D$-$\av{\psi}/\psi_D$ plane in (b),
where $\psi_m$ is the midpoint value.  
In (a) and (b),  the capillary condensation curves  (bold lines) are 
calculated from 1d  profiles, where  
points for five  $\tau/\tau_D$ 
are those along the $z$ axis with  
 $(x,y)=(D,D)$ and $(D,0)$ 
in the final two-phase states  in our simulation  (see Figs.4 and 5). 
Bulk coexistence curve is in broken line.  
In (b) our phase separation process  is illustrated by arrows, 
where the total order parameter is conserved. 
}
\end{center}
\end{figure}

Figure 1  displays  typical 1d  profiles  
 of  $\psi(z)$ from Eq.(2.29) and  
$\omega_{\rm{loc}}(z)$ in Eq.(2.22) in the range $0<z<D/2$, 
which  will be needed  
to explain our simulation results.  Here we set   
$(\tau/\tau_D, \psi_\infty/\psi_D, \mu_\infty/\mu_D)
=(-2, -1.14, -14.1)$ (top), 
 $(-20, -1.85,-30.8)$ (middle), and $(-20, -1.81,-17.8)$ (bottom), where 
$\mu_\infty$ is measured in   units   of      
\be 
\mu_D= T_c/D^3\psi_D\propto D^{\nu/\beta-3}.
\en  
 Salient features in  Fig.1 are as follows.   
(i) In Fig.1(a), $\psi$ is positive 
  in the whole region with $<\psi>/\psi_D=1.20$.  
(ii) In Figs.1(b) and (b'), 
the fluid is on the capillary condensation   line, 
where we give  two equilibrium 
profiles B and C with    the same  $\Omega$. 
Here, B represents an adsorption-dominated state with 
$\psi_m \sim \psi_{\rm cx}$  and  $<\psi>/\psi_D=1.99$, 
while for C  the film center 
is occupied by the  disfavored phase  
with $\psi_m \sim -\psi_{\rm cx}$ and $<\psi>/\psi_D=0.187$. 
In the right panel (b'),   
the integral of $\omega_{\rm{loc}}(z)$ in the region $0<z<D/2$ is the  same for B and C.  The enclosed two regions have the same 
area $24.7$ in units of $T_c/D^2$, 
which is close to  the surface tension $\sigma=29.2T_c/D^2$ 
 at this $\tau$.  In addition, $\Pi \sim -\sigma/D$ for B 
and $\Pi \cong 0$ for C \cite{comment2}. 
(iii) In Figs.1(c) and (c'), 
the parameters are those slightly below 
the  capillary condensation   line (in Fig.2 below). 
Here, there are three solutions with  the common   $\tau$ and $\mu_\infty$,  
but    $<\psi>/\psi_D$ is $  2.03$ for (D), 
 0.028 for (E), and  1.20 for (F)  (see Fig.3 below). 
If we perform simulation in contact with a reservoir 
with  these  $\tau$ and $\mu_\infty$, the profile D is realized at long times. 
In the very early stage of our simulation, 
the dynamics is one-dimensional and the profile 
 F is    approached  after quenching from  A in Fig.1(a) 
(see Fig.7 below).

In  Fig.2,  we  show the capillary condensation  line  (CCL) from 1d 
profiles  located on the 
left of the bulk coexistence line  in the 
$\tau$-$\mu_\infty$ plane. In our  previous paper \cite{Oka}, 
 the corresponding phase diagram was displayed   in the 
$\tau$-$\psi_\infty$ plane. 
The  discontinuities of the physical quantities  
across CCL  increase with increasing $|\tau|$ 
vanishing  at a film critical point. At this  film criticality, 
$\tau$, $\psi_\infty$, and $\mu_\infty$  are  calculated as  
\be  
\bigg (\frac{\tau}{\tau_D}, 
\frac{\psi_\infty}{\psi_D}, \frac{\mu_\infty}{\mu_D}\bigg)= 
(-3.14, -1.27, -16.3),
\en    
where we also have $\av{\psi}/\psi_D=0.989$ and $\psi_m/\psi_D=-0.173$.   
Hereafter, the   chemical potential $\mu_\infty$ 
on this CCL   will  be  written as 
$ \mu_{\rm cx}^{\rm ca}(\tau)$. 
Our  numerically calculated  CCL  is well  
fitted to the linear form, 
\be 
\mu_{\rm cx}^{\rm ca}(\tau) 
/\mu_D+ 16.3  \cong 0.86 (\tau/\tau_D + 3.14).
\en  

In  Fig.1(b'), 
 the two areas enclosed by the two curves 
of $\omega_{\rm loc}D^3/T_c$  are the same ($\sim \sigma D^2/T_c$). 
Thus, for $|\tau|/\tau_D \gg 1$,  
the surface tension  $\sigma$ 
and  the free energy difference per unit area 
$-2\mu_{\rm cx}^{\rm ca} \psi_{\rm cx}D$ are of the same order. 
See the sentences below Eq.(2.30) 
and the  explanation of Fig.1(b').  
 For $|\tau|\gg 
\tau_D$, it  follows the relation,   
\be 
  \mu_{\rm cx}^{\rm ca} \sim   - \sigma/\psi_{\rm cx}D 
\sim - |\tau/\tau_D|^{2\nu-\beta} \mu_D. 
\en 
Since   $2\nu-\beta \cong 0.94$, 
the   theoretical formula (2.34)   
is consistent with the  numerical formula (2.33).
Note that Eq.(2.34)  is equivalent to 
the Kelvin equation known for the gas-liquid  transition 
in pores  \cite{Evansreview,Gelb}.

We have  already presented a special case of  three  1d profiles 
 in Fig.1(c)  for $(\tau/\tau_D,\mu_\infty/\mu_D)= 
(-20,-17.8)$. In  Fig.3, we show 
isothermal curves in the  
$\mu_\infty$-$\av{\psi}$ plane,
which  are calculated from  1d profiles with   
$\tau/\tau_D= -2$, $-10$,  and $-20$. The relation between  
$\mu_\infty$ and $\av{\psi}$  is monotonic for 
$\tau/\tau_D\ge -3.14$ (above the film critical temperature), 
while    it exhibits 
a van der Waals loop for $\tau/\tau_D<-3.14$  with  
 three 1d states   
in a window range $\mu_{\infty 1}< \mu_\infty< \mu_{\infty 2}$ \cite{comment3}. Here, $\mu_{\infty 1}$ 
and $\mu_{\infty 2}$   
coincide at the film criticality. 
The isothems  consist of  stable 
and unstable parts characterized by the sign of  the 
 film susceptibility defined by 
\be 
\chi_{\rm film}= T_c 
(\p \av{\psi}/\p \mu_\infty)_\tau. 
\en   
In Fig,3, points A,B,C,D,E, and F correspond to the curves in Fig.1.
Dotted parts of the  two curves of $\tau/\tau_D= -10$ and $-20$ 
 are not  stable in the presence of a mass current 
from  a reservoir with common $\mu_\infty$.   

Previously, 
some authors  \cite{p,Binder-Landau} calculated  the  stable   parts  
of isotherms  of  
 the average density in the film versus  the chemical potential.  
In our  local functional theory,  
the three 1d profiles can be calculated since  a unique 
profile follows for any given set of 
 $\tau$ and $\av{\psi}$. 
 In equilibrium  fluctuation theory of  films 
 \cite{Evansreview},  $\chi_{\rm film}$  
is proportional to the  variance 
of the order parameter fluctuations, so its negativity 
indicates thermodynamic instability.

Furthermore,  Fig.4 gives  the phase diagrams      in the 
 $\tau$-${\psi_m}$ and $\tau$-${\av{\psi}}$ planes.  
Bold lines represent the capillary condensation curve from 1d profiles 
as in Fig.2.  
In steady two-phase states in our simulation, 
$\psi_m$ and $\av{\psi}$ depend on $x$ and $y$, so points 
for five $\tau$ represent 
$\psi_m(x,y) =\psi(x,y,D/2)$ and  $\av{\psi}(x,y)$ 
  with   $(x,y)=(D,D)$ and $(D,0)$  (see Figs.5 and 6). 
The former  line passes  through   a   domain  
of the phase disfavored by the walls   
and the latter  through the favored phase only. 
Phase diagrams similar to Fig.4(b) have been obtained 
in experiments of the capillary condensation in porous media \cite{Gelb}.

\begin{table}
\caption{  Values of  
$\mu_{\infty}/\mu_D$ from our simulation in   a 
finite $2D\times 2D\times D$  system  for five $\tau/\tau_D$. 
The corresponding values of  $\mu_{\rm cx}^{\rm ca}/\mu_D$ 
on the CCL from 1d profiles are also shown.  
}
\begin{tabular}{|c||c|c|c|c|c|} 
\hline
$\tau/\tau_D$ & $-8.5$ & $-10$ & $-13.3$  & $-16.6$ & $-20$\\
\hline
$\mu_{\infty}/\mu_D$ (finite) &-25.8&-27.8 & -32.2 &-36.6 &-41.4 \\
\hline 
$\mu_{\rm cx}^{\rm ca}/\mu_D$ (1d)&-20.6&-21.8 & -24.9 &-27.7 &-30.8 \\
\hline
\end{tabular}
\end{table}

\section{Phase separation dynamics }

We performed simulation 
of phase separation in a $L\times L\times D$ cell 
with $L=2D $ imposing 
 the periodic boundary condition  along the $x$ and $y$ axes. 
In this section, we describe phase separation realized 
for  deep quenching. 
However, it was not realized 
for shallow quenching $(|\tau|/\tau_D \ls 7$),  
for which $\chi_{\rm film}$ in Eq.(2.35) is positive.

\begin{figure}[t]
\begin{center}
\includegraphics[scale=0.4]{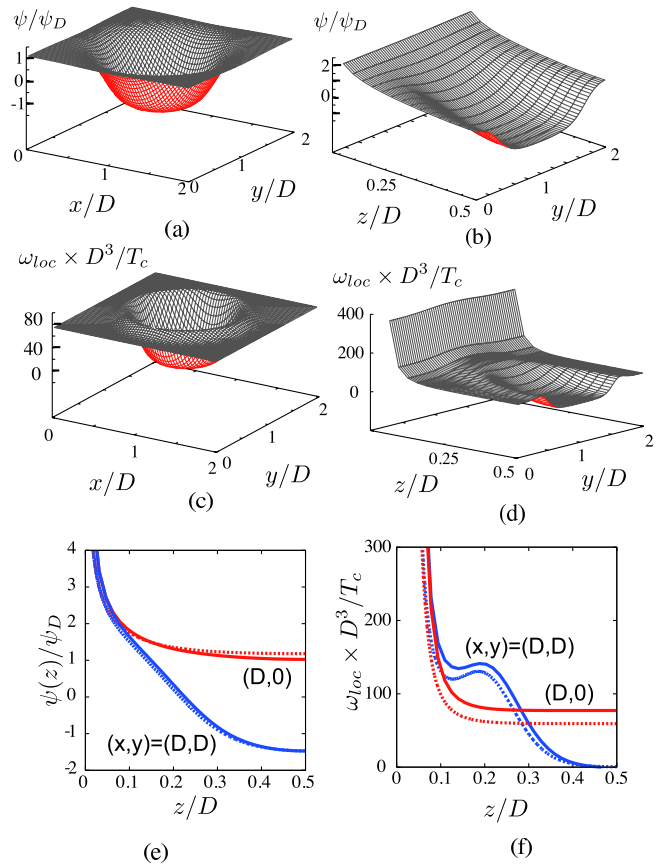}
\caption{(Color online) Equilibrium two-phase state   at  
 $\tau/\tau_D=-10$ in a $2D\times 2D \times D$ system: 
$\psi/\psi_D$ (top) 
and $\omega_{\rm loc}D^3/T_c$ (middle) 
in the $xy$ plane at $z=D/2$ (left) and 
in the $yz$ plane at $x=D$ (right).
Bottom: $\psi/\psi_D$ (left) and $\omega_{\rm loc}D^3/T_c$ (right) 
along the $z$ axis for   $(x,y)=(D,D)$ (blue bold line) 
and $(D,0)$ (red bold line), while  
 dotted lines represent  1d profiles from Eq.(2.29). 
}
\end{center}
\end{figure}

\begin{figure}[t]
\begin{center}
\includegraphics[scale=0.4]{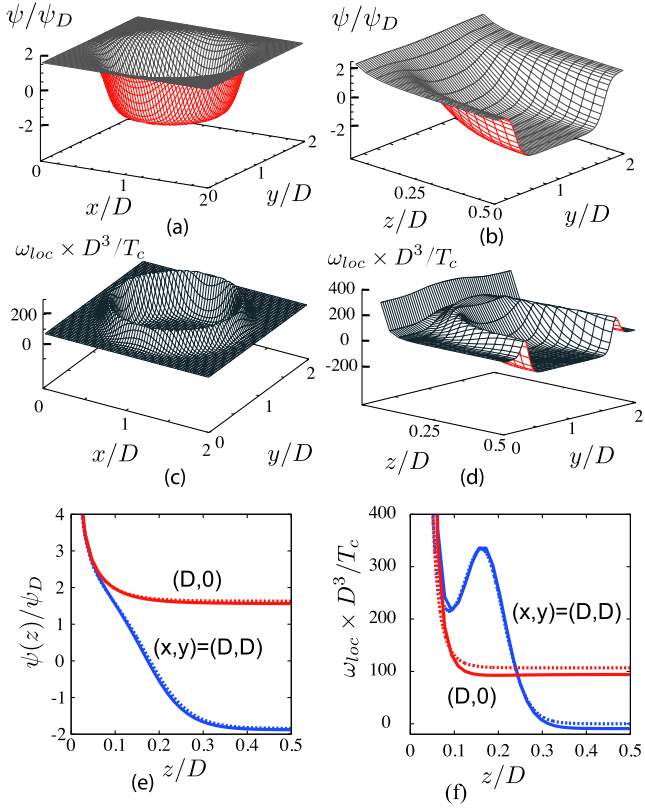}
\caption{(Color online) Equilibrium two-phase state   at  
 $\tau/\tau_D=-20$ in a $2D\times 2D \times D$ system: 
$\psi/\psi_D$ (top) and $\omega_{\rm loc}D^3/T_c$ (middle) 
in the $xy$ plane at $z=D/2$ (left) and 
in the $yz$ plane at $x=D$ (right).
Bottom:  $\psi/\psi_D$ (left) and $\omega_{\rm loc}D^3/T_c$ (right) 
along the $z$ axis for   $(x,y)=(D, D)$ (blue bold line) 
and $(D,0)$ (red bold line), which are closer 
to the  1d profiles (dotted lines) from Eq.(2.29) than in Fig.5. 
}
\end{center}
\end{figure}

\setcounter{equation}{0}
\subsection{Dynamic equations and simulation method}

Supposing   an incompressible  
   fluid binary mixture  
with a homogeneous  temperature, we 
use the model H equations \cite{H77,Kawasaki-Ohta,Onukibook}. 
The order parameter  $\psi$ is a conserved variable governed by     
\be 
\frac{\p \psi}{\p t} =-\nabla\cdot(\psi {\bi v}) +  
\lambda \nabla^2 \frac{\delta F}{\delta\psi}, 
\en  
where $\lambda$ is the kinetic coefficient 
and the functional derivative ${\delta F}/{\delta \psi}$ 
may be calculated  from Eq.(2.4) 
with the aid of Eqs.(2.10), (2.15), and (2.16) 
outside and inside CX. 
We neglect the random source term originally present in critical dynamics
\cite{H77,Onukibook}, because we treat the deviations much larger than 
the thermal fluctuations. 
The velocity field $\bi v$ satisfies $\nabla\cdot{\bi v}=0$   
and vanishes   at  $z=0$ and $D$.   
In the Stokes approximation\cite{Kawasaki-Ohta},  
 $\bi v$  is determined  by  
\be 
\bar{\eta} \nabla^2{\bi v}=\nabla p_0 + \psi \nabla  (\delta F/\delta\psi),  
\en  
where $\bar{\eta}$ is the shear viscosity and the role of 
a pressure $p_0$ is  to ensure  
$\nabla\cdot{\bi v}=0$. See Appendix for the expression 
of the stress tensor in near-critical fluids and the  derivation of Eq.(3.2). 
  
The kinetic coefficients  $\lambda$ and $\bar{\eta}$ should be treated as 
renormalized ones \cite{H77,Kawasaki-Ohta,Onukibook} 
(see the last sentence of Subsec.IIA).  
In  the vicinity of 
the bulk coexistence curve, 
 $\lambda$ may be approximated  by   
\be 
\lambda=   \chi_{\rm cx}D_\xi /T_c, 
\en 
where $ \chi_{\rm cx}$ is the susceptibility on CX 
in Eq.(2.13) and $D_\xi$ is 
 the mutual  diffusion constant of the Stokes form, 
\be 
D_\xi= T_c /6\pi \bar{\eta} \xi ,
\en 
with  $\xi=\xi_0' |\tau|^{-\nu}$ being  the correlation length  on CX.
In our simulation, 
$|\psi|$ is of order $ \psi_{\rm cx}$ 
at $z=D/2$ (see Figs.5 and 6 below), 
which supports Eqs.(3.3) and (3.4). 
The viscosity $\bar\eta$ exhibits a very weak 
critical singularity    and  may be treated as 
a constant  independent of $\tau$.

In this paper, we also performed simulation 
for model B without the  hydrodynamic interaction \cite{H77},  
where  $\psi$ obeys the diffusive equation,  
\be 
\frac{\p \psi}{\p t} = \lambda \nabla^2 \frac{\delta F}{\delta\psi}.  
\en  
The kinetic coefficient 
 $\lambda$ is assumed to be given by Eqs.(3.3) and (3.4) 
as in the model H case. Then, comparing 
the results from the two models,  we can examine    the 
role   of the hydrodynamic interaction in phase separation. 
Model  B has  been used 
to investigate  surface-directed phase separation in binary alloys 
\cite{Das}.

In integrating Eqs.(3.1) and (3.5), 
the mesh length was  $\Delta x=D/32$ and 
the time interval width was 
 $\Delta t= 2\times 10^{-6} t_0$. 
The initial state was  the 1d profile A 
in Fig.1(a) at $\tau/\tau_D= -2$ 
 with   small random numbers ($\sim 10^{-4}$) 
superimposed  at the mesh points. 
At $t=0$, we decreased $\tau$ 
to a final reduced temperature. 
 For $t>0$,  there was  no  mass exchange  
between  the film and the reservoir 
so that the total order 
parameter $\int d{\bi r} \psi$ was  fixed   at $1.20\psi_D DL^2$. 
We will measure  time after quenching  in units of 
\be 
t_0= D^2/D_\xi,
\en  
which is the mutual diffusion time in  the film  
assumed to be much longer than the thermal diffusion time. 
We note that  the natural time unit 
in bulk phase separation has been 
the order parameter relaxation time $t_\xi= \xi^2/D_\xi$ 
\cite{Kawasaki-Ohta,Siggia,Onukibook}. Here, 
$t_0/t_\xi= (D/\xi)^2= R_\xi^{2} |\tau/\tau_D|^{2\nu}\sim 
100$ for $|\tau/\tau_D|\sim 10$.

\begin{figure}[t]
\begin{center}
\includegraphics[scale=0.17]{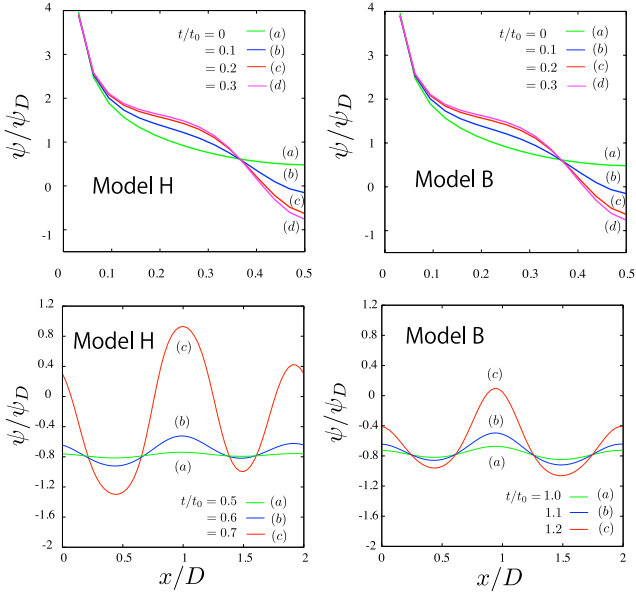}
\caption{(Color online)  Early-stage time-evolution  
 of  $\psi/\psi_D$ for model H (left) 
and model B (right) after quenching from 
the profile A in Fig.1(a) to $\tau/\tau_D=-20$. 
Shown  are the profiles 
along the $z$ axis 
for $(x,y)=(D,D)$ at $t/t_0=0$,  
0.1, 0.2, and 0.3  (top) and along the $x$ axis  
for $(y,z)=(1.5D,0.5D)$ (bottom) 
at later times.  In the top panels,  the profile F in Fig.1(c) 
is approached  without noticeable differences  
between  the two  models. In the bottom panels, 
fluctuations in the $xy$ plane appear 
and  coarsening is much  quickened for model H 
than for model B, where $t/t_0=$ 
0.5, 0.6, and 0.7 
for model H  and   1.0, 1.1, and 1.2 for model B.  
}
\end{center}
\end{figure}

\begin{figure}[htbp]
\begin{center}
\includegraphics[scale=0.17]{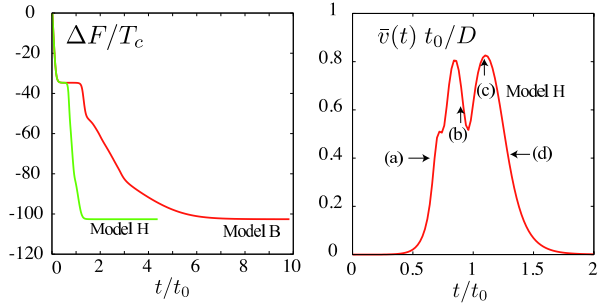}
\caption{(Color online)  Time-evolution  
 of  the normalized free energy decrease $\Delta F(t) /T_c$ for model H  
and model B (left) and $\bar{v}(t) t_0/D$ for model H (right) 
 after quenching from 
the profile A in Fig.1(a). 
At points (a), (b), (c), and (d) (right) 
snapshots of $\psi$ and $\bi v$ will be given in Fig.9. }
\end{center}
\end{figure}

\begin{figure*}[t]
\begin{center}
\includegraphics[scale=0.245]{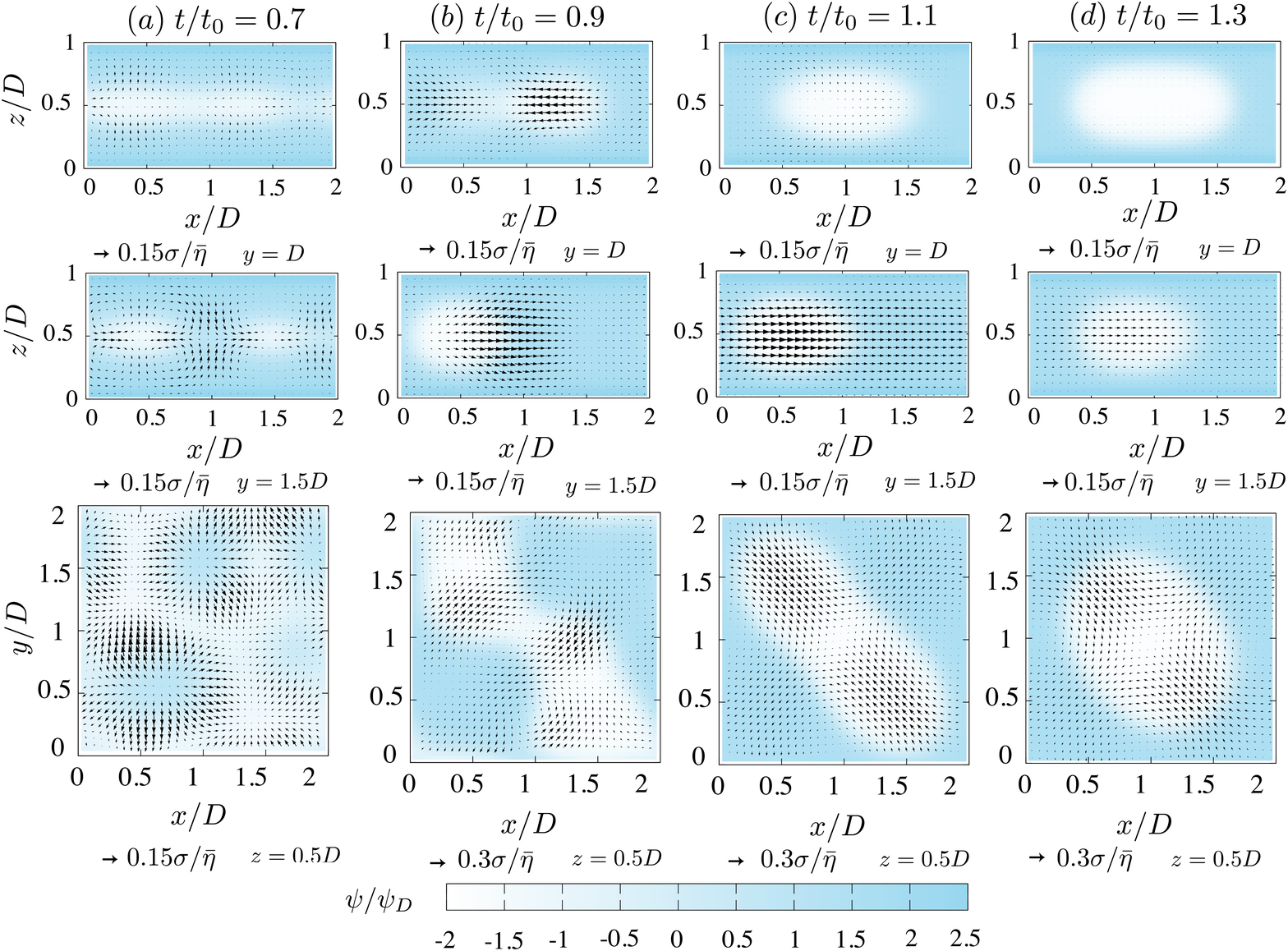}
\caption{(Color online) Cross-sectional velocity field ${\bi v}$ (arrows) 
and order parameter $\psi$ (in gradation 
according the color bar)  in phase separation for $\tau/\tau_D=-20$ 
in model H, 
where   $t/t_0$ is equal to 
(a) 0.7, (b) 0.9, (c) 1.1, and (d) 1.3 after quenching. 
Displayed are  $(v_x,v_z)$ and $\psi$ in the $xz$  plane at  $y=D$ (top),  
those   at  $y=1.5D$ (middle),   and  $(v_x,v_y)$ and $\psi$ 
 in the $xy$ plane at $z=D/2$ (bottom). 
Arrows below panels 
indicate  typical magnitudes of the velocities, 
where $\sigma$ is the surface tension and $\bar\eta$ is the 
shear viscosity. Final state is given in Fig.6. 
 }
\end{center}
\end{figure*}

\subsection{Steady  two-phase states }

For sufficiently deep  quenching, we realized 
  phase separation to find   a   steady  two-phase 
state at long times both for   model H and model B. 
 We could also  calculate this   final state   
more accurately from the following relaxation-type   equation, 
\be 
\frac{\p \psi}{\p t} = 
-L_0  \bigg[ \frac{\delta F}{\delta\psi}-
\AV{\frac{\delta F}{\delta\psi}}_t \bigg], 
\en 
where $L_0$ is a constant and 
$\av{{\delta F}/{\delta\psi}}_t$ is the 
space average of ${\delta F}{\delta\psi}$ at time $t$. 
Because of its simplicity, we integrated Eq.(3.7) with 
a fine   mesh length  of  $\Delta x=D/64$.  
The data points in Figs.2 and 4 
and the snapshots  in Figs.5 and 6  are those  
from the steady states of Eq.(3.7).

In Fig.4, the  deviations of $\psi_m$ and $\av{\psi}$ 
from those on CCL  are surprisingly   small,  
though the lateral dimension $L$ is only $2D$. 
However, in Table 1, the final two-phase values of $\mu_\infty$ 
considerably deviate from those on  CCL. 
This may be ascribed to the relatively small size of the 
susceptibility $\chi_{\rm cx}$ 
for these cases. That is, 
if we set $\chi_{\rm cx}= A_\chi^{-1}  T_c\psi_D/\mu_D$ on CX, we 
obtain $A_\chi=120$ and  283 for $\tau/\tau_D=-10$ and $-20$, 
respectively.
Here, if we multiply 
the deviation  of ${\psi_m}/\psi_D$ or $\av{\psi}/\psi_D$ 
by of $A_\chi$, we obtain that of   
 $\mu_\infty/\mu_D$.

In Figs.5 and 6, we   
display the final  profiles of $\psi$  
and $\omega_{\rm loc}$  
in the $xy$ plane at $z=D/2$ (left) and 
in the $zy$ plane at $x=D$ (right) for   
 $\tau/\tau_D=-10$ and $-20$. 
In these cases, $\xi$ on CX is $0.078D$ and $0.051D$, respectively, 
which is of the order of the interface thickness. 
Displayed in the  bottom panels  are 1d profiles  
of $\psi$  and $\omega_{\rm loc}$  along the $z$ axis 
for  the two lateral points $(x,y)=(D,D)$ and $(D,0)$. 
These profiles  are rather close to 
the  1d profiles from Eq.(2.29) 
in accord with Fig.4.

\subsection{Time evolution}

Both for model H and model B, early-stage 
 time-evolution proceeds   as follows. 
 Just after quenching, 
$\psi$ changes  only along the $z$ axis 
to approach the 1d profile at the final $\tau$ 
with fixed $\av{\psi}$ (see   Fig.3). 
If this  1d profile 
satisfies the instability condition  $\chi_{\rm film}<0$, 
it follows  3d  spinodal decomposition. 
On the other hand, if it  satisfies the stability condition   
 $\chi_{\rm film}>0$,  
it remains stationary  in  simulation without   thermal noise.

 Figure 7 displays  $\psi$ 
after quenching to  $\tau/\tau_D=-20$. In 
the top panels, it  is plotted  along the $z$ axis 
with  $(x,y)=(D,D)$  at $t/t_0= 0,0.1,0.2,$ and 0.3. 
The  velocity field nearly  vanishes 
for  model H, so there is almost no 
difference between the results of these  models. 
However, in the bottom  panels,   the 
1d profile becomes  unstable with respect to the fluctuations  
varying in the $xy$ plane for $t/t_0 \gs 0.5$. 
The   velocity field grows   gradually for model H.  
In this second time range, 
the domain  formation is much  quicker for model H than for model B.

In the left panel of Fig.8, we show  
the free energy decrease $\Delta F= F(t)- F(+0)$  at $\tau/\tau_D=-20$ 
as a function of $t$ for model H and model B. Here, 
 $F(t)$ is the total bulk free energy in Eq.(2.4) 
 with   $F(+0)$ being   
its value just  after quenching. 
Its decrease  is accelerated with development of 
the fluctuations in the $xy$ plane. 
The coarsening is slower for model B than for model H 
by about 5 times. 
In the right panel of Fig.8, we show 
the characteristic velocity  amplitude ${\bar v}(t)$ for model H,  
which is  defined by 
\be 
{\bar v}(t)^2 = \int d{\bi r} |{\bi v}|^2 / DL^2. 
\en 
For  $\tau/\tau_D=-20$, 
 ${\bar v}(t)$ 
 is equal to 0.021, 0.106, and 0.485 
at $t/t_0=0.5$, 0.6, and 0.7, respectively,  
increasing up to $0.8 $, in units of $D/t_0$.

In Fig.9, we show late stage snapshots of $\psi$ and $\bi v$ 
in the $xz$ plane (top) and in the $xy$ plane (middle) for $\tau/\tau_D=-20$ 
for  model H, 
where   $t/t_0$ is equal to 
(a) 0.7, (b) 0.9, (c) 1.1, and (d) 1.3 after quenching. 
In (a) we can see a network-like domain of the disfavored phase. 
In (b) three domains can be seen, where the middle one 
is being absorbed into the bottom  one, soon resulting in  two domains 
at $t/t_0\sim 1.0$.  This process gives 
rise to a  dip in  ${\bar v}(t)$ in Fig.8(b) since  
these two domains are considerably apart. 
In (c) and (d), furthermore, coalescence of these two domains is 
 taking place.    The arrows  below the panels 
indicate  the 
typical  velocity, $0.15 \sigma/\bar{\eta}$ 
or   $0.3 \sigma/\bar{\eta}$, 
where $\sigma$ is the surface tension and $\bar\eta$ is the 
shear viscosity.  
Note  that  
the typical velocity in the 
late-stage bulk spinodal decomposition 
is given by  $v_c= 
0.1\sigma/\bar{\eta}$ \cite{Siggia,Onukibook}, which follows 
from the stress balance $\sigma/R \sim 6\pi {\bar \eta} v_c/R$ 
with $R \sim v_c t$ being the typical domain length. 
In our case, from Eq.(2.17) these velocities are related as  
\be 
 D/t_0=T_c/6\pi{\bar\eta}\xi D= 
(\xi/6\pi A_\sigma D)(\sigma/{\bar\eta}).
\en 
Thus, $ D/t_0= 0.036 \sigma/\bar{\eta}$  for 
$\tau/\tau_D=-20$ in Fig.9.

\section{Summary and remarks}

In summary, we  have examined  the phase separation 
in  a  near-critical  binary mixture between symmetric parallel plates  
in the strong adsorption regime around the capillary condensation line (CCL). 
Using model H and model B, 
simulation has been performed in  a $2D\times 2D\times D$ cell.  
We summarize our main results.\\ 
(i) In Sec.II, we have presented  the singular 
free energy with the gradient part 
outside and inside the bulk coexistence curve. Applying it 
to near-critical fluids between parallel plates, 
typical 1d profiles have been given in Fig.1.  
CCL  has been plotted in the $\tau$-$\mu_\infty$ plane in Fig.2. 
The points for the  steady two-phase states 
from our simulation   are located in the left side of CCL. 
In Fig.3, we have also found the  van der Waals loop of  isothermal curves   
in the  $\av{\psi}$-$\mu_\infty$ plane, where 
 $\tau$ is smaller than the film critical value.   
The phase diagrams have been plotted in the  $\tau$-$\psi_m$ 
 and $\tau$-$\av{\psi}$ planes  in  3d (bulk) and 2d (film) in Fig.4. 
The Kelvin relation (2.34) has also been obtained, 
since the osmotic pressure 
  $\Pi$ is of order $-\sigma/D$ right below CCL \cite{comment2}.\\ 
(ii) In Sec.III,   
we have first displayed the cross-sectional 
profiles of $\psi$ and $\omega_{\rm loc}$ 
in steady two-phase states in Figs.5 and 6. 
The profiles along $z$ axis for 
$(x,y)=(D,D)$ and $(D,0)$  closely resemble the corresponding  
1d profiles. For quenching to $\tau/\tau_D=-20$, 
we have examined time-evolution of $\psi$.   
It  occurs only along the $z$ axis 
in the very  early stage in the top panels of Fig.7, 
where there is no difference between the results of model  H and model B. 
Subsequently, inhomogeneities appear in the $xy$ plane. 
The free energy decrease $\Delta F(t)=F(t)- F(+0)$ and the 
typical velocity amplitude ${\bar v}(t)$ defined in Eq.(3.8) 
have been plotted in Fig.8. 
The velocity field considerably quicken the 
interface formation and the coarsening  for  model H  
than for model B. 
Profiles of $\psi$ and $\bi v$ 
in the late stage coarsening 
due to the flow have  been presented in Fig.9, 
where the domain coalescence 
can be seen and the maximum velocity is of order $0.1\sigma/{\bar \eta}$ 
as in  bulk spinodal decomposition \cite{Siggia}.

We make some  remarks.\\
1) In the static part of our theory, 
we neglect the thermal fluctuations 
varying in the lateral directions 
with wavelengths longer than $D$. 
Thus this 2d transition exhibits  mean-field  behavior. 
In fact, the curves of $\psi_m$ vs $\tau$ 
and $\av{\psi}$ vs $\tau$ are parabolic near the film criticality 
in Fig.4.\\ 
2)  In our simulation,  
we soon have only two or three domains 
in the cell as in the bottom panels of Fig.9.  
The lateral dimension $L=2D$  in this paper is too short 
to investigate  the domain growth law in the $xy$ plane. 
Simulation with larger $L/D$ should be performed in future.\\  
3) From the van der Waals loop of the isothermal curves 
 in the $\av{\psi}$-$\mu_\infty$ 
plane in Fig.3, we may predict how  phase separation proceeds after quenching. 
We have examined phase separation via spinodal decomposition. 
However, in real experiments, phase separation may occur via nucleation    
for  metastable 1d profiles.  Note that  
hysteretic behavior has been observed 
in phase-separating   fluids in pores 
and has not been well explained  \cite{Evansreview,Gelb,Goldburg,Wil}.
\\ 
4) In  a number of experiments and simulations 
of surface-directed phase separation \cite{Araki,Bates,Puri},  
 composition waves  along the $z$ axis have been 
 observed near the wall  in the early stage. 
 In  these cases, the degree of  adsorption 
  has  changed appreciably  upon quenching.   
In  the  strong adsorption regime in this paper,   1d dynamics   
occurs  in the initial stage, but there are no 
composition waves as in the top panels of Fig.7.  
\\ 
5) The static part of this work is 
applicable to any  Ising-like near-critical 
systems   and can readily be generalized to 
$n$-component spin systems. 
In the dynamics, we have used  model H with 
 a homogeneous temperature and incompressible flows. 
On the other hand, in one-component near-critical fluids,  
the latent heat 
released or absorbed at the interfaces  gives rise to    
significant  hydrodynamic flow because of the enhanced 
isobaric thermal expansion \cite{Onukibook,Teshi}. 
Also promising in future should be 
extension of this work to near-critical 
fluids in   porous media.\\

\begin{acknowledgments}
This work was supported by Grant-in-Aid 
for Scientific Research  from the Ministry of Education, 
Culture,  Sports, Science and Technology of Japan. 
S. Y. was supported by the Japan Society for Promotion of Science.
A.O. would like to thank Sanjay Puri for informative correspondence. 
\end{acknowledgments}

\vspace{2mm}
\noindent{\bf Appendix: Stress tensor in near-critical fluids  }\\
\setcounter{equation}{0}
\renewcommand{\theequation}{A\arabic{equation}}
 
In near-critical fluids, we treat slow 
 flows with  low   Reynolds numbers. 
The total (reversible) stress tensor is given by 
$p_0\delta_{ij}+\Pi_{\psi ij}$, where  $p_0$ is   nearly 
homogeneous  throughout 
the film and the reservoir.  The $\Pi_{\psi ij} $ is the  stress tensor 
due to the composition deviation\cite{Onukibook},  
\be 
\Pi_{\psi ij} = p_\psi \delta_{ij}   +C (\nabla_i\psi) (\nabla_j  \psi), 
\en 
where $\nabla_i  = \p /\p x_i$. 
The diagonal part $p_\psi$ is written as 
\bea 
p_\psi&=& \psi \mu- f -\frac{1}{2} (C+C'\psi)|\nabla\psi|^2 -C \psi\nabla^2\psi\nonumber\\ 
&=&\psi(\delta F/\delta \psi )- f- C |\nabla\psi|^2/2, 
\ena 
where $C'= \p C/\p \phi$.  The second  part 
 in Eq.(A1) contains off-diagonal components 
relevant  for curved interfaces.  
In deriving Eq.(3.2), we use the relation, 
\be 
\sum_j \nabla_j \Pi_{\psi ij} = \psi \nabla_i (\delta F/\delta \psi).
\en 


\begin{references}

\bibitem{Evansreview}
R. Evans and U. M. B. Marconi, 
J. Chem. Phys. {\bf 86}, 7138 (1987);  
 R. Evans, J. Phys.: Condens. Matter {\bf 2}, 8989 (1990).

\bibitem{Gelb} 
L.D. Gelb, K.E. Gubbins, R. Radhakrishnan, and 
M. Sliwinska-Bartkowiak, Rep. Prog. Phys. {\bf 62}, 1573 (1999).

\bibitem{Binder} K. Binder, D.  Landau, and M. M$\ddot{\rm u}$ller, 
J. Stat. Phys. {\bf 110}, 1411 (2003); 
M. M$\ddot{\rm u}$ller  and K. Binder, J. Phys.: Condens. Matter 17, S333
(2005); K.Binder,  J. Horbach, R. Vink,  and A. De Virgiliis, 
Soft Matter, {\bf 4}, 1555 (2008). 


\bibitem{Cahn} J. W. Cahn, 
J. Chem. Phys. {\bf 66} 3667 (1977).

\bibitem{Binderreview} 
K. Binder, in {\it Phase Transitions and Critical Phenomena}, 
C. Domb and J. L. Lebowitz, 
eds. (Academic, London, 1983), Vol. 8, p. 1.



\bibitem{PG} P.G. de Gennes, 
Rev. Mod. Phys. {\bf 57}, 827 (1985). 


\bibitem{Bonn} 
D. Bonn and D. Ross, Rep. Prog. Phys. {\bf 64}, 1085 (2001).

\bibitem{Lawreview} 
B. M. Law, Prog. Surf. Sci. {\bf 66}, 159 (2001). 

\bibitem{Rudnick}
J. Rudnick and D. Jasnow, 
Phys. Rev. Lett. {\bf 48}, 1059 (1982);
ibid. {\bf 49}, 1595  (1982)  


\bibitem{Liu} A. J. Liu and M. E. Fisher, 
Phys. Rev. A {\bf 40}, 7202 (1989).


\bibitem{Nakanishi}
M. E. Fisher and H. Nakanishi,
J. Chem. Phys. {\bf 75}, 5857 (1981); 
H. Nakanishi and M. E. Fisher,
J. Chem. Phys. {\bf 78}, 3279 (1983).


\bibitem{Evans-Marconi} 
R. Evans,   U. M. B. Marconi, 
 P.  Tarazona,  J. Chem. Soc.,Faraday.
Trans.  {\bf  82}, 1763 (1986);   
P. Tarazona, U.M.B  Marconi, R.  Evans, 
Mol. Phys. {\bf 60}, 573 (1987);
 

\bibitem{p} 
B. K. Peterson, K. E. Gubbins, G. S. Heffelfinger, U. Marini, 
B. Marconi,  and F. van Swol, 
J. Chem. Phys. {\bf 88}, 6487 (1988).
	
\bibitem{Binder-Landau} 
K. Binder and D. P. Landau, J. Chem. Phys. {\bf 96}, 1444 (1992).


\bibitem{chemical} 
For binary mixtures, we use  
the chemical potential difference 
$\mu_1-\mu_2$ betwen the two components. 
To be precise, $\mu_\infty$ 
in the text is  the deviation 
of the chemical potential difference 
$\mu(c_\infty,T)- \mu(c_c,T)$ 
around the critical composition $c_c$. 

\bibitem{Evans-Anna}
A. Macio\l ek, A. Drzewi\'nski,and R. Evans, 
Phys. Rev. E {\bf 64},  056137 (2001). 

\bibitem{Singh} S. K. Singh, A. K. Saha, and J. K. Singh, 
 J. Phys. Chem. B {\bf 114}, 4283 (2010).


\bibitem{Oka} R. Okamoto and A. Onuki,   
J. Chem. Phys. {\bf 136}, 114704 (2012).

\bibitem{Yang} 
M. E. Fisher and H. Au-Yamg, 
Physica {\bf 101}A,  255 (1980).

  
\bibitem{Upton} 
M. E. Fisher and P. J. Upton, 
Phys. Rev. Lett. {\bf 65}, 3405 (1990); 
Z. Borjan and P. J. Upton, {\it ibid.} 
{\bf  81}, 4911 (1998); {\it ibid.} 
 {\bf 101}, 125702 (2008). 


\bibitem{Casimir} 
A. Gambassi, A. Macio\l ek, C. Hertlein, U. Nellen, 
L. Helden, C. Bechinger, and S. Dietrich, 
Phys. Rev. E {\bf 80}, 061143 (2009).  

\bibitem{Tsori} S. Samin and Y. Tsori, 
EPL {\bf 95},  36002 (2011). 

\bibitem{Okamoto} 
R. Okamoto and A. Onuki, Phys. Rev. E {\bf 84}, 051401 (2011).
\bibitem{H77} P.C. Hohenberg and B.I.  Halperin,  { Rev. Mod. Phys.} {\bf 49}, 435 (1977).

\bibitem{Onukibook} 
A. Onuki, {\it Phase Transition Dynamics} (Cambridge University Press, Cambridge, 2002).
\bibitem{Goldburg} 
M. C. Goh, W. I. Goldburg, and C. M. Knobler, 
Phys. Rev. Lett. {\bf 58}, 1008 (1987);  
A. P. Y. Wong,  S. B. Kim,  W. I. Goldburg,  and M. H. W. Chan, 
{\it ibid.} {\bf 70}, 954 (1993). 

\bibitem{Wil} 
S. B. Dierker and P. Wiltzius, 
Phys. Rev. Lett. {\bf 58}, 1865 (1987). 

\bibitem{Liu-Durian} 
A. J. Liu, D. J. Durian, E. Herbolzheimer, and S. A. Safran, 
Phys. Rev. Lett.{\bf 65}, 1897 (1990). 

\bibitem{Bates} 
R. A. L. Jones, L. J. Norton, E. J. Kramer, F. S. Bates, and
P. Wiltzius, Phys. Rev. Lett. {\bf 66}, 1326 (1991).

\bibitem{Tanakareview} 
H. Tanaka, Phys. Rev. Lett. {\bf 70}, 53 (1993); 
{\it ibid.} {\bf 70}, 2770 (1993);  
J. Phys.: Condens. Matter {\bf 13}, 4637 (2001).

\bibitem{Das} 
S. K. Das, S. Puri, J. Horbach, and K. Binder, Phys. Rev. E {\bf 72},
061603 (2005).



\bibitem{Yeomans} M. R. Swift, W. R. Osborn, and J.M. Yeomans, 
Phys. Rev. Lett. {\bf 75}, 830 (1995).

\bibitem{Araki}  H. Tanaka and T. Araki, 
Europhys. Lett. {\bf  51} 154 (2000).

\bibitem{Puri} 
P. K. Jaiswal, S. Puri, and S. K. Das, Phys. Rev. E {\bf 85}, 051137 (2012); 
EPL, {\bf 97},  16005  (2012). 



\bibitem{Monson}
F. Porcheron and P. A. Monson, Langmuir {\bf 21}, 3179 (2005). 






  

\bibitem{Sc69} P. Schofield,  Phys. Rev. Lett. {\bf 22}, 606 (1969); 
 P. Schofield, J.D. Lister, and J.T. Ho,  ibid. {\bf 23}, 1098 (1969). 



 \bibitem{comment0} 
Reliable values of 
$R_\chi$ and $R_\xi$ in the literature 
are 4.9 and 1.9, respectively\cite{Oka,Onukibook}.
The correlation
length  and the susceptibility  on CX 
are considerably underestimated in our theory. 




 \bibitem{comment2} 
The scaled quantity  ${\cal A} \equiv - D^3\Pi /T_c>0$  
 is one of the 
Casimir amplitudes depending on  
$\tau/\tau_D$ and $\psi_\infty/\psi_D$   \cite{Oka}. Its  
value at the bulk criticality 
is ${\cal A}_{\rm critical} = 0.558$.  
For $|\tau/\tau_D|\gg 1$, 
 we  obtain ${\cal A} \cong A_{\rm os} |\tau/\tau_D|^{2\nu}\gg 1$ 
with  $A_{\rm os}\cong 2$ 
just below  CCL (for (B) in Fig.1) 
and ${\cal A}\cong 0$ just above CCL (for (C) in Fig.1). 



\bibitem{comment3} In Fig.13 of Ref.\cite{Oka},  
we already found a cubic relation 
between $\psi_m$ (instead of $\av{\psi}$) 
and $\psi_\infty$ (instead of  $\mu_\infty$) 
near the film criticality for $\tau/\tau_D>-3.14$, 
which is  a mean-field result 
due to neglect of   2d thermal fluctuations with sizes longer than $D$. 


\bibitem{Kawasaki-Ohta}
K. Kawasaki, Prog. Theor. Phys. {\bf 57}, 826 (1977); K.
Kawasaki and T. Ohta, {\it ibid}. {\bf 59}, 362 (1978).

\bibitem{Siggia}  E. D. Siggia, 
Phys. Rev. A {\bf 20}, 595 (1979). 


\bibitem{Teshi} R. Teshigawara and A.  Onuki, 
Phys. Rev. E 82, 021603 (2010); {\it ibid.} {\bf 84}, 041602 (2011).


\end{references}
\end{document}